\documentclass[prd,showpacs]{revtex4}
\begin{document}
\title{Spherical Casimir effect for a massive scalar field on the three dimensional ball}
\author{Andrea Erdas}
\email{aerdas@loyola.edu}
\affiliation{Department of Physics, Loyola University Maryland, 4501 North Charles Street,
Baltimore, Maryland 21210, USA}
\begin {abstract} 
The zeta function regularization technique is used to study the Casimir effect
for a scalar field of mass $m$ satisfying Dirichlet boundary
conditions on a spherical surface of radius $a$. In the case of large scalar mass, $ma\gg1$, simple analytic expressions are obtained for the 
zeta function and Casimir energy of the scalar field when it is confined inside the spherical surface, and when it is confined outside the spherical surface.
In both cases the Casimir energy is exact up to order $a^{-2}m^{-1}$ and contains the expected divergencies, which can be eliminated using
the well established renormalization procedure for the spherical Casimir effect.
The case of a scalar field present in both the interior and exterior region is also examined and,
for $ma\gg 1$, the zeta function, the Casimir energy, and the Casimir force are obtained. The obtained Casimir energy and force are exact up to order $a^{-2}m^{-1}$ 
and $a^{-3}m^{-1}$ respectively. In this scenario both energy and force are finite and do not need to be renormalized, and the force is found to produce an outward pressure on the spherical surface.
\end {abstract}
\maketitle
\section{Introduction}
\label{1i}
The electromagnetic Casimir effect was first predicted theoretically by H. G. B. Casimir \cite{Casimir:1948dh} in 1948, when he showed that an attractive force 
exists between two electrically neutral, parallel conducting plates in vacuum. 
Boyer predicted the repulsive Casimir force some time later, when he discovered that a perfectly conducting, neutral spherical surface in vacuum modifies
the vacuum energy of the electromagnetic field in such a way that the spherical surface is subject to an outward pressure \cite{Boyer:1968uf}. 
Experimental confirmation of the Casimir effect came more than fifty years ago by Sparnaay \cite{Sparnaay:1958wg}, and 
many improved experimental observations have been reported throughout the years \cite{Bordag:2001qi,Bordag:2009zz}.

Since their discovery, Casimir forces have been found to have many applications from nanotechnology to string theory, and a large effort has gone 
into studying the generalization of the Casimir effect to quantum fields other than the electromagnetic field: fermions
were first considered by Johnson \cite{Johnson:1975} then investigated by many others, and bosons and other scalar fields
have also been investigated extensively \cite{Bordag:2001qi}. 

It is well known that Casimir forces
are very sensitive to the boundary conditions of the quantum fields at the plates. In the case of scalar fields, Dirichlet and Neuman
boundary conditions are most commonly used, in the case of fermion fields or other fields with spin \cite{Ambjorn:1981xw},
bag boundary conditions are used.
In this manuscript we investigate a scalar field that obeys Dirichlet boundary conditions on a spherical surface of radius $a$. 

Massive or massless scalar fields appear in many areas of physics from
the Higgs field in the Standard Model, to the dilaton field that breaks the conformal symmetry in 
string theory, to the Ginzburg-Landau scalar field in superconductivity, etc. 
The spherical Casimir effect for massless \cite{Bender:1994zr,Cognola:1999zx} or massive \cite{Bordag:1996ma,Kirsten:2000ad,
Saharian:2000mw,BezerradeMello:2000mb}
scalar fields in $(3+1)$ or  $(D+1)$ dimensions has been studied in vacuum and at finite temperature \cite{Balian:1977qr,Teo:2013bza}
using the Green's function method \cite{Bender:1994zr,Saharian:2000mw} or the zeta function technique 
\cite{Cognola:1999zx,Bordag:1996ma,Kirsten:2000ad,BezerradeMello:2000mb} to calculate the Casimir energy. 
These authors however, are only able to obtain the Casimir energy for large scalar mass as an infinite sum of 
hypergeometric  functions.
In this manuscript I use the zeta function technique
to study the spherical Casimir effect for a scalar field of mass $m$, and obtain 
simple analytic forms for the zeta function and Casimir energy when the scalar field is confined inside or outside the 
spherical surface, in the case of large scalar mass ($ma\gg 1$). In both cases the Casimir energy is found to be divergent,
as expected \cite{Bordag:1996ma,Kirsten:2000ad}. I also obtain simple expressions for the large mass Casimir energy 
and force on the spherical surface in the case of a scalar field present both inside and outside the spherical surface. 
The energy and force obtained for this scenario are finite.

In Section \ref{2i}, I describe the model and, for the case of a scalar field confined 
inside the spherical surface, obtain the zeta function $\zeta^{int}(s)$ using the Debye uniform asymptotic expansion of the 
modified Bessel functions.
In Section \ref{3i}, I find a simple expression for $\zeta^{int}(s)$ in the large mass limit.
In Section \ref{4i}, I use the large mass limit of $\zeta^{int}(s)$ to calculate the Casimir energy for a scalar field confined inside the spherical surface,
in the case of $ma\gg 1$. I also obtain, using $\zeta^{int}(s)$ from Section \ref{3i}, the large mass limit of the zeta function and Casimir energy
in the case of a scalar field confined outside the spherical surface. Finally I study the case of a scalar field present both inside and outside
the spherical surface, and find very simple analytic expressions for the Casimir energy and force on the spherical shell, when $ma\gg 1$. 
A summary and discussion of my results are presented in Section \ref{5i}. 
\section{ Zeta function inside a spherical surface}
\label{2i}

In 3-dimensional space the equation of motion of a scalar field, $\phi(x)=\phi({\bf x})e^{-i\omega t}$, is the Klein-Gordon equation
$$\left(-\Delta +m^2\right)\phi({\bf x})=\omega^2\phi({\bf x}),
$$
where $m$ is the scalar field mass. Using spherical coordinates, this equation becomes
\begin{equation}
{1\over r^2}\left[-{\partial\over \partial r}\left(r^2{\partial\over \partial r}\right)+{\hat L}^2+r^2m^2\right]\phi(r,\theta,\varphi)
=\omega^2\phi(r,\theta,\varphi),
\label{1}
\end{equation}
where $\hat {\bf L}$ is the angular momentum operator. After a separation of variables
$$\phi(r,\theta,\varphi)=g(r)Y_{l m}(\theta,\varphi),$$
the radial part of Eq. (\ref{1}) is found to be
\begin{equation}
{1\over r^2}\left[-{d\over d r}\left(r^2{d\over d r}\right)+l(l+1)+r^2m^2\right]g(r)
=\omega^2g(r).
\label{2}
\end{equation}
A complete set of solutions of Eq. (\ref{2}), finite at the origin, is
$$g(r)=r^{-1/2}J_{l+1/2}(\bar{\omega} r),$$
where $J_{l+1/2}(z)$ are Bessel functions of the first kind and $\bar{\omega}^2=\omega^2-m^2$. Once we impose Dirichlet 
boundary conditions on a spherical surface of radius $a$
$$g(a)=0,$$
we find
$$\bar{\omega} ={w_{l,n}\over a},$$
where $w_{l,n}$ is the $n$-th zero of $J_{l+1/2}(z)$ and $n=1,2,\cdots$. The energy eigenvalues are found immediately
$$\omega_{l,n}^2=\left({w_{l,n}\over a}\right)^2+m^2,$$
and, when the scalar field is confined inside the spherical surface, the zeta function is given by
\begin{equation}
\zeta^{int}(s)=\sum_{l=0}^\infty\sum_{n=1}^\infty(2l+1)(\omega_{l,n})^{-2s},
\label{3}
\end{equation}
where $2l+1$ is the degeneracy of the eigenmodes of angular momentum $l$. 

Since ${\partial
\over \partial k}\ln J_{l+1/2}(k a)$ has simple poles at $k={w_{l,n}\over a}$, I can write Eq. (\ref{3}) in the form of 
a contour integral \cite{Bordag:1995gs,Bordag:1995gm}
$$\zeta^{int}(s)=\sum_{l=0}^\infty(2l+1)\oint_\gamma{dk\over 2\pi i}(k^2+m^2)^{-s}{\partial
\over \partial k}\ln J_{l+1/2}(k a)
$$
where the closed contour $\gamma$ runs counterclockwise, contains the whole positive $k$-axis and, with it, 
all of the ${w_{l,n}\over a}$ for $l\ge 0$ and $n\ge 1$. Next I rotate the integration contour to the imaginary axis
and obtain
\begin{equation}
\zeta^{int}(s)={\sin(\pi s)\over \pi}\sum_{l=0}^\infty2\nu \int_m^\infty{dk}(k^2-m^2)^{-s}{\partial
\over \partial k}\ln\left[k^{-\nu}I_{\nu}(k a)\right],
\label{4}\end{equation}
where $\nu=l+{1\over 2}$, $I_\nu(z)$ is a modified Bessel function of the first kind, and the added factor $k^{-\nu}$ inside 
the logarithm does not change the result, since no additional pole is enclosed. A simple change of the integration variable
allows me to rewrite Eq. (\ref{4}) as
\begin{equation}
\zeta^{int}(s)={2\sin(\pi s)\over \pi}a^{2s}\sum_{l=0}^\infty \nu \int_{am/\nu}^\infty{dz}(\nu^2z^2-a^2m^2)^{-s}{d
\over d z}\ln\left[z^{-\nu}I_{\nu}(\nu z)\right],
\label{5}\end{equation}
and to exploit the Debye uniform asymptotic expansion of the modified Bessel functions \cite{Bordag:1996ma}
\begin{equation}I_{\nu}(\nu z)\sim{1\over\sqrt{2\pi\nu}}{e^{\nu \eta}\over (1+z^2)^{1/4}}\sum_{k=0}^\infty{u_k(t)\over \nu^k}
\label{6}\end{equation}
where 
$$\eta = \sqrt{1+z^2}+\ln{z\over 1+\sqrt{1+z^2}},\;\;\;\;\;\;\;\;t={1\over\sqrt{1+z^2}}
$$
and $u_k(t)$ is defined recursively by
$$u_0(t)=1,\;\;\;\;\;\;\;\;u_k(t)={t^2(1-t^2)\over 2}u'_{k-1}(t)+{1\over 8}\int_0^t(1-5\tau^2)u_{k-1}(\tau)d\tau .
$$
I use Eq. (\ref{6}) and find
\begin{equation}{d\over d z}\ln\left[z^{-\nu}I_{\nu}(\nu z)\right]\sim{\nu\over z}\left(\sqrt{1+z^2}-1\right)-{1\over 2}{z\over z^2+1}-
\sum_{i=1}^N{zt^3D'_i(t)\over\nu^i}
\label{7}\end{equation}
where the $D_i(t)$ are defined through
$$\ln\left[\sum_{k=0}^\infty{u_k(t)\over \nu^k}\right]=\sum_{i=1}^\infty{D_i(t)\over\nu^i},
$$
and are polynomials of degree $3i$
$$D_i(t)=\sum_{j=0}^i x_{ij}t^{i+2j},
$$
while the coefficients $x_{ij}$ can be easily calculated with a simple computer program.
It is clear that, as $N$ grows, the right side of Eq. (\ref{7}) becomes a more accurate approximation of the left side of (\ref{7}). Using Eq. (\ref{7}),
I write the following approximate expression of the zeta function
\begin{equation}
\zeta^{int}(s)\sim\sum_{i=-1}^NA_i(s),
\label{8}\end{equation}
with
$$A_{-1}(s)={2\sin(\pi s)\over \pi}a^{2s}\sum_{l=0}^\infty \nu^2 \int_{am/\nu}^\infty(\nu^2z^2-a^2m^2)^{-s}
{\sqrt{z^2+1}-1\over z}{dz},
$$
$$A_{0}(s)=-{\sin(\pi s)\over \pi}a^{2s}\sum_{l=0}^\infty \nu \int_{am/\nu}^\infty(\nu^2z^2-a^2m^2)^{-s}{z\over z^2+1}{dz},
$$
and, for $i\ge 1$
$$A_{i}(s)=-{2\sin(\pi s)\over \pi}a^{2s}\sum_{l=0}^\infty \nu \int_{am/\nu}^\infty(\nu^2z^2-a^2m^2)^{-s}{z\over (z^2+1)^{3/2}}{D_i'(t)\over \nu^i}{dz}
.$$
Equation (\ref{8}) displays the same feature as Eq. (\ref{7}): as $N$ grows the sum on the right side
becomes a more accurate approximation of the zeta function. 
\section{Zeta function in the large mass limit}
\label{3i}

In this section I evaluate the $A_i(s)$ of Eq. (\ref{8}) in the limit $ma\gg 1$. 
When $z\gg 1$, I can write
$${\sqrt{z^2+1}-1\over z}\simeq 1-{1\over z} +{1\over 2z^2}+{\cal O}(z^{-4}),
$$
and therefore, in the large mass limit, I find
\begin{equation}
A_{-1}(s)\simeq {2\sin(\pi s)\over \pi}a^{2s}\sum_{l=0}^\infty \nu^2 \int_{am/\nu}^\infty(\nu^2z^2-a^2m^2)^{-s}
\left[1-{1\over z} +{1\over 2z^2}+{\cal O}(z^{-4})\right]{dz}.
\label{9}\end{equation}
Similarly
$${z\over z^2+1}\simeq z^{-1}\left[1- {1\over z^2}+{\cal O}(z^{-4})\right],
$$
$${z\over (z^2+1)^{3/2}}\simeq z^{-2}\left[1- {3\over 2z^2}+{\cal O}(z^{-4})\right],
$$
$$t={1\over\sqrt{1+z^2}}\simeq z^{-1}\left[1- {1\over 2z^2}+{\cal O}(z^{-4})\right],
$$
when $z\gg 1$, and thus $A_0(s)$ and $A_i(s)$ become
\begin{equation}
A_{0}(s)\simeq-{\sin(\pi s)\over \pi}a^{2s}\sum_{l=0}^\infty \nu \int_{am/\nu}^\infty(\nu^2z^2-a^2m^2)^{-s}z^{-1}\left[1- {1\over z^2}
+{\cal O}(z^{-4})\right]{dz},
\label{10}\end{equation}
\begin{equation}
A_{i}(s)=-{2\sin(\pi s)\over \pi}a^{2s}\sum_{l=0}^\infty \nu^{1-i} \int_{am/\nu}^\infty(\nu^2z^2-a^2m^2)^{-s}\sum_{j=0}^i {x_{ij}\over z^{1+i+2j}}
(i+2j)\left[1- {2+i+2j\over 2z^2}+{\cal O}(z^{-4})\right]{dz}. 
\label{11}\end{equation}
After I change the integration variable from $z$ to $y=\nu^2z^2-a^2m^2$ in the integrals of Eqs. (\ref{9}) - (\ref{11}), use
$$z^{-s}={1\over \Gamma(s)}\int_0^\infty \alpha^{s-1}e^{-z\alpha}d\alpha,
$$
and integrate over the new variable $y$, I obtain
$$A_{-1}(s)= {\sin(\pi s)\over \pi}\Gamma(1-s)\sum_{l=0}^\infty \left(a\over \nu\right)^{2s} \int_0^\infty e^{-\alpha m^2a^2}\alpha^{s-3/2}
\left[{1\over\sqrt\pi}-\nu\alpha^{1/2}+{\nu^2\over\sqrt\pi}\alpha+{\cal O}(\alpha^{2})\right]{d\alpha},
$$
$$A_{0}(s)= -{\sin(\pi s)\over2 \pi}\Gamma(1-s)\sum_{l=0}^\infty \left(a\over \nu\right)^{2s} \int_0^\infty e^{-\alpha m^2a^2}\alpha^{s-1}
\left[1-\nu^2\alpha+{\cal O}(\alpha^{2})\right]{d\alpha},
$$
and, for $i\ge 1$
$$A_{i}(s)= -{2\sin(\pi s)\over \pi}\Gamma(1-s)\sum_{l=0}^\infty \sum_{j=0}^i \left(a\over \nu\right)^{2s}{\nu^{2j}x_{ij}\over \Gamma({i\over 2}+j)} \int_0^\infty e^{-\alpha m^2a^2}\alpha^{s-1+i/2+j}
\left[1-\nu^2\alpha+{\cal O}(\alpha^{2})\right]{d\alpha}.
$$
The integrals over $\alpha$ are done easily, and I find
\begin{equation}
A_{-1}(s)= {\sin(\pi s)\over \pi a}{\Gamma(1-s)\over \left(a m\right)^{2s}}
\left[{am\over\sqrt\pi}\zeta_H(2s,{\textstyle\frac{1}{2}})\Gamma(s-{\textstyle\frac{1}{2}})-\zeta_H(2s-1,{\textstyle\frac{1}{2}})\Gamma(s)+
{\zeta_H(2s-2,{\textstyle\frac{1}{2}})\over\sqrt\pi am}\Gamma(s+{\textstyle\frac{1}{2}})+{\cal O}\left({\textstyle\frac{1}{a^3m^3}}\right)\right],
\label{12}\end{equation}
\begin{equation}
A_{0}(s)= - {\sin(\pi s)\over 2\pi a}{\Gamma(1-s)\over \left(a m\right)^{2s}}
\left[\zeta_H(2s,{\textstyle\frac{1}{2}})\Gamma(s)-{\zeta_H(2s-2,{\textstyle\frac{1}{2}})\over a^2m^2}\Gamma(s+1)+{\cal O}\left({\textstyle\frac{1}{a^4m^4}}\right)\right],
\label{13}\end{equation}
and, for $i\ge 1$
$$A_{i}(s) = -{2\sin(\pi s)\over \pi a}{\Gamma(1-s)\over \left(a m\right)^{2s}}\sum_{j=0}^i {x_{ij}\over \Gamma({i\over 2}+j)(am)^{i+2j}}
\biggl[\zeta_H(2s-2j,{\textstyle\frac{1}{2}})\Gamma(s+{\textstyle\frac{i}{2}}+j)
$$
\begin{equation}
\left.-{\zeta_H(2s-2j-2,{\textstyle\frac{1}{2}})\over a^2m^2}\Gamma(s+{\textstyle\frac{i}{2}}+j+1)+{\cal O}\left({\textstyle\frac{1}{a^4m^4}}\right)\right],
\quad\quad\quad\quad
\label{14}\end{equation}
where
$$
\zeta_H(s,z)=\sum_{l=0}^\infty (l+z)^{-s}
$$
is the Hurwitz zeta function. 

\section{Casimir energy and force in the large mass limit}
\label{4i}

The Casimir energy for a massive scalar field confined inside a spherical surface of radius $a$, is given by
$$ E^{int}=\frac{1}{2}\lim_{\epsilon \rightarrow 0}\zeta^{int}(\epsilon - {\textstyle\frac{1}{2}}),
$$
where $\zeta^{int}(s)$ is given by Eq. (\ref{8}), and therefore I obtain the large mass limit of $ E^{int}$ using Eqs. (\ref{12}) - (\ref{14})
for the large mass limits of the $A_i(s)$.  I find
\begin{equation}
aA_{-1}(\epsilon - {\textstyle\frac{1}{2}})= {a^2m^2\over 48 \pi}
\left[\frac{1}{\epsilon} -3 -{12\zeta'_R(2)\over \pi^2}+2\gamma_E-2\ln\left({am\over 2\pi}\right)\right]
+ {7\over 1920 \pi}\left[\frac{1}{\epsilon} +12\ln2 -14 +{1680\zeta'_R(-3)\over \pi^2}-14\ln\left(am\right)\right]
\label{15}\end{equation}
for ${\epsilon \rightarrow 0}$, where $\zeta_R(z)$ is the Riemann zeta function of number theory and $\gamma_E=0.5772$
is the Euler Mascheroni constant, and where I neglected all terms of order $1\over (am)^{n}$ with $n \ge 2$, 
since $am\gg 1$. Similarly, I find
\begin{equation}
aA_{0}(\epsilon - {\textstyle\frac{1}{2}})=- {am\over 48 }
+ {7\over 3840 \pi am},
\label{16}\end{equation}
\begin{equation}
aA_{1}(\epsilon - {\textstyle\frac{1}{2}})= {1\over 192\pi }
\left[\frac{1}{\epsilon} -4 -{12\zeta'_R(2)\over \pi^2}+2\gamma_E-2\ln\left({am\over 2\pi}\right)\right],
\label{17}\end{equation}
\begin{equation}
aA_{2}(\epsilon - {\textstyle\frac{1}{2}})= {1\over 384am},
\label{18}\end{equation}
where I used $x_{10}=\frac{1}{8}$, $x_{20}=\frac{1}{16}$, and neglected all terms of order $1\over (am)^{n}$ with $n \ge 2$.
Notice that, since the contribution of $aA_{i}(\epsilon - {\textstyle\frac{1}{2}})$ is of order $1\over (am)^{i-1}$, I do not need to consider any $A_i$ with 
$i> 2$, and therefore the exact large mass limit of $aE^{int}$ to order $\frac{1}{am}$ is
\begin{equation}
aE^{int}=\frac{1}{2}\lim_{\epsilon \rightarrow 0}\sum_{i=-1}^2aA_i(\epsilon - {\textstyle\frac{1}{2}}),
\label{19}\end{equation}
where the $aA_i(\epsilon - {\textstyle\frac{1}{2}})$ are given by Eqs. (\ref{15}) - (\ref{18}). Notice also that, as $s \rightarrow -\frac{1}{2}$, 
the expected divergencies  \cite{Bordag:1996ma} only appear in $A_{-1}$ and $A_1$, due to the 
factors of $\Gamma(s-\frac{1}{2})$ and $\Gamma(s+\frac{1}{2})$ present inside Eqs. (\ref{12}) and (\ref{14}). All the other $A_i$ are free of 
divergencies  as $s \rightarrow -\frac{1}{2}$, since
$\Gamma(s+\frac{i}{2}+j)$ is finite when $i>1$ and $j\ge0$.

If the scalar field is confined outside the spherical surface, the zeta function is  \cite{Bordag:1996ma}
\begin{equation}
\zeta^{ext}(s)\sim\sum_{i=-1}^N(-1)^iA_i(s),
\label{20}\end{equation}
and can be used to calculate $E^{ext}$, the Casimir energy for the exterior region. The large mass limit of $aE^{ext}$, exact to order $\frac{1}{am}$, is
\begin{equation}
aE^{ext}=\frac{1}{2}\lim_{\epsilon \rightarrow 0}\sum_{i=-1}^2(-1)^iaA_i(\epsilon - {\textstyle\frac{1}{2}}),
\label{21}\end{equation}
with the $aA_i(\epsilon - {\textstyle\frac{1}{2}})$ given by Eqs. (\ref{15}) - (\ref{18}). Since $A_{-1}(\epsilon - {\textstyle\frac{1}{2}})$
and $A_{1}(\epsilon - {\textstyle\frac{1}{2}})$ contain divergent terms as ${\epsilon \rightarrow 0}$, the Casimir energy for the 
exterior region is also divergent, as expected. The renormalization of the divergent Casimir energies $E^{int}$ and $E^{ext}$ can be carried out using the procedure 
outlined, for example, in Ref. \cite{Bordag:1996ma}.

Finally, I discuss the situation where the scalar field is present in both the interior and exterior regions. In this case the 
Casimir energy is
$$E=E^{int}+E^{ext}$$
and, using Eqs. (\ref{19}) and (\ref{21}), I find a finite value for the large mass limit of $E$
\begin{equation}
E=-{m\over 48}+\left(10+\frac{7}{\pi}\right){1\over 3840 a^2m}.
\label{22}\end{equation}
Notice that, as $m\rightarrow \infty$, the Casimir energy $E\rightarrow -\infty$.
The Casimir force $F$ on the spherical surface of radius $a$ is given by
$$F=-{\partial E \over \partial a},
$$
and I find a repulsive force
\begin{equation}
F=\left(10+\frac{7}{\pi}\right){1\over 1920 a^3m},
\label{23}\end{equation}
indicating an outward pressure on the spherical surface, that vanishes as $m\rightarrow\infty$. The large mass limit Casimir energy $E$ and force $F$ that
I find in Eqs. (\ref{22}) and (\ref{23}), are exact to order $\frac{1}{a^2m}$ and $\frac{1}{a^3m}$ respectively.
\section{Discussion and conclusions}
\label{5i}
In this manuscript I used the zeta function regularization technique to study the spherical Casimir effect
of a massive scalar field in $(3+1)$ dimensions. I analyzed three scenarios: a scalar field confined inside a spherical surface, a scalar field confined outside the spherical surface, 
and a scalar field present inside and outside the spherical surface at the same time. In all cases Dirichlet boundary conditions were imposed on the sphere of
radius $a$. I obtained two expressions 
of the zeta function in the large mass limit, one valid inside the sphere and one valid outside, which are exact to order $1\over a^2 m^2$, 
and used them to obtain the large mass limit of the Casimir energy inside (\ref{19}) and outside the sphere (\ref{21}), exact to order  $1\over a^2 m$. These Casimir energies contain divergencies,
as I expected, and can be renormalized following the renormalization procedure described in Ref. \cite{Bordag:1996ma}.

I also studied the case of a scalar field present both inside and outside the spherical surface, and obtained the large mass limit of the Casimir energy
(\ref{22}) and force (\ref{23}) in this case. Both quantities are finite and thus do not need to be renormalized, and are exact to order 
$\frac{1}{a^2m}$ and $\frac{1}{a^3m}$ respectively. 

For a scalar field with mass $m_H\sim 100$ GeV, such as the Higgs, I find that any spherical surface of radius $a\ge a_H$, with $a_H\sim 2$ Fm, 
abundantly satisfies 
the large mass condition, since $m_Ha_H=10^3$. In this scenario, I find that the Casimir force on the spherical surface is $F\le F_H$, 
where $F_H\sim 3.2\times 10^2$ eV$\cdot$fm$^{-1}$ is obtained by using $m_H$ and $a_H$ in Eq.(\ref{23}).


\end{document}